\documentclass[10pt,conference]{IEEEtran}
\IEEEoverridecommandlockouts
\usepackage{cite}
\usepackage{amsmath,amssymb,amsfonts}
\usepackage{algorithmic}
\usepackage{graphicx}
\usepackage{balance}
\usepackage{textcomp}
\usepackage{comment}
\usepackage{xcolor}
\usepackage{xspace}
\usepackage{booktabs}
\usepackage{array}
\usepackage{multirow}
\usepackage{url}
\usepackage{tabularx}
\def\BibTeX{{\rm B\kern-.05em{\sc i\kern-.025em b}\kern-.08em
    T\kern-.1667em\lower.7ex\hbox{E}\kern-.125emX}}

\usepackage{caption}

\newcommand{\appname}{\textsc{IssueLighter}\xspace}

\begin{document}

\title{\appname: An Issue Type-Aware LLM Agent for Issue Report Validity Assessment}

\author{
\IEEEauthorblockN{Jiawei Ran,
Yuxia Zhang\textsuperscript{\dag},
Ziyang Zhao,
Hui Liu}
\IEEEauthorblockA{
\textit{School of Computer Science \& Technology,}\\
\textit{Beijing Institute of Technology,}\\
Beijing, China\\
\{jiaweiran, yuxiazh, ziyangzhao, Liuhui08\}@bit.edu.cn}
\thanks{\textsuperscript{\dag}Corresponding author.}
}

\maketitle

\begin{abstract}
Invalid issue reports are prevalent in open source software (OSS) projects, imposing substantial burdens on maintainers who already face a high risk of burnout. The widespread adoption of large language models (LLMs) in software development has further exacerbated this problem. 
Prior research has primarily focused on identifying invalid reports for specific issue categories (e.g., bug and wontfix), leaving the broader landscape of OSS issue types largely unexplored and limiting the generalizability of existing approaches. Moreover, existing methods provide little support for explaining or validating their predictions, reducing their practical usefulness for maintainers.
With the rise of LLM-based agents, it remains unclear whether a comprehensive understanding of the causes of invalidity across all issue types can improve validity assessment performance while producing well-grounded rationales. To this end, we first manually annotated the validity status of 520 issue reports from three OSS projects, i.e., pandas, scikit-learn, and mypy, resulting in 257 invalid reports and 263 valid ones. Then, we conducted a thematic analysis on these invalid reports and identified 13 causes of invalidity that are closely specific to issue types. Based on these insights, we propose \appname, an agent-based framework with a type-aware validity assessment strategy that dynamically executes commands. Experimental results show that \appname consistently and substantially outperforms all baseline methods, with an average F1 improvement of over 20 percentage points. Ablation studies confirm that both the type-aware strategy and the taxonomy of causes of invalidity contribute to the performance gains. Qualitative analysis further demonstrates that \appname generates rationales grounded in accurately identified causes of invalidity and supported by evidence. Moreover, leveraging these rationales, we challenged maintainers’ original decisions in four disputed cases, and all received approval, highlighting the practical value of our approach.
\end{abstract}


\section{Introduction}
In the maintenance of open source software (OSS), issue reports serve as the primary communication channel between users and development teams~\cite{bertram2010communication}. However, a significant number of submitted issue reports are invalid, i.e., the claimed problem or request is itself unfounded or unreasonable, and are ultimately marked as not-planned, etc. While these reports do not lead to any project changes, maintainers still invest substantial time reading, diagnosing, and responding to them. This cost is far from negligible. A study of 279 GitHub projects revealed that developers take an average of five months to close a wontfix issue~\cite{panichella2021won}. Research on deep learning compilers further showed that false-positive reports (accounting for 41.7\% of all reports) require comparable handling time to real issue reports~\cite{huang2025false}. This phenomenon is also prevalent in software developed in companies~\cite{guo2010characterizing, laiq2024industrial}. Maintainers are already well-known for their burnout in the numerous requests they receive, e.g., issue reports, code reviews, and support requests~\cite{Raman2020Stress}. Recently, as LLMs make it effortless to generate plausible issue reports, OSS projects are experiencing an increasing influx of AI-generated submissions, many of which require substantial manual verification despite ultimately being invalid or non-actionable~\cite{afroz2026aislop}\footnote {\url{https://lkml.org/lkml/2026/5/17/896}}. 
These findings highlight the urgent need for automated approaches to reduce the effort spent on invalid reports.

To address the challenge of invalid reports, the research community has proposed various automated approaches, including feature engineering-based classifiers~\cite{zanetti2013categorizing, fan2018chaff}, traditional machine learning methods~\cite{panichella2021won}, and deep learning models~\cite{he2020deep,meng2023bug}. Although these approaches achieve promising performance, they only produce a binary prediction (valid or invalid). In practice, however, maintainers also need to understand why an issue is considered invalid to facilitate effective communication with reporters and subsequent issue handling. Existing studies have investigated the causes of invalidity, but almost exclusively for issue reports labeled as bugs~\cite{sun2011bug,erfani2014works,rahman2020some,huang2025false} and wontfix~\cite{wang2020my,panichella2021won}. Building on such taxonomies, Huang et al.~\cite{huang2025false} employed LLMs to identify invalid bug reports and generate corresponding rationales. However, existing studies remain limited to a small subset of issue categories, leaving the causes of invalidity across the broader landscape of OSS issue types largely unexplored. Moreover, their rationales lack supporting evidence for validity assessment, making it difficult for maintainers to verify and act upon the results.

The rise of LLM agent mechanisms offers a new direction for overcoming this limitation. Unlike traditional static approaches, agents can actively invoke external tools, forming a ``think-act-observe'' loop~\cite{yao2022react,schick2023toolformer,yang2023intercode}. This provides the technical foundation for enabling executable validation and leveraging additional resources. To this end, this paper aims to move the needle by characterizing root causes of all the issue types, then guiding the LLM agent to identify invalid issue reports with precise rationales.    

We first empirically investigate the root causes of invalid issue reports across all issue types, with particular attention to type-specific causes. Specifically, we ask \textbf{RQ1: What are the root causes of invalid issue reports across different issue types?} We conducted a thematic analysis on 257 invalid reports from three popular OSS projects (i.e., pandas, scikit-learn, and mypy). We identified 13 causes of invalidity across six issue types, i.e., bug, enhancement, documentation, continuous integration (CI), installation, and spam. Furthermore, we found that different issue types exhibit fundamentally different patterns of causes of invalidity. 

Based on the findings from RQ1, we propose \appname, an agent-based framework with a validity assessment strategy tailored to each issue type that dynamically executes commands. Specifically, we ask \textbf{RQ2: What is the performance of \appname compared to the baselines?} We answer this question by evaluating \appname and the other baselines on our well-annotated issue reports. The results show that \appname consistently outperforms baseline methods across three LLMs, achieving 96.2\% accuracy and 96.0\% F1-score on ChatGPT in particular. We also conducted a qualitative analysis of the rationales given by \appname. We found that its rationales align closely with the causes of invalidity we annotated in RQ1 and are grounded in empirical evidence rather than textual inference. Furthermore, based on the rationale given by \appname, we challenged maintainers’ original decisions in four disputed cases, all of which were approved, highlighting the practical value of our approach.

The main contributions of this paper are as follows:
\begin{itemize}
    \item A comprehensive taxonomy of causes of invalidity across six issue types.
    \item \appname, an automated framework for issue report validity assessment that achieves substantial performance gains over baselines.
    \item A publicly available benchmark of 520 manually annotated issue reports with validity labels and root causes of invalidity. 
\end{itemize}

\section{Related Work}\label{related_work}

\subsection{Characterization of Invalid Issue Reports}

Understanding the causes of invalid issue reports is key to automated detection. Sun et al. manually analyzed over 600 invalid bug reports and found that testing errors, misunderstandings of functionality, and faults introduced by external libraries were the primary causes~\cite{sun2011bug}. Focusing specifically on non-reproducible reports, Joorabchi et al. identified six categories of root causes, with inter-bug dependencies and environmental differences being the most prevalent~\cite{erfani2014works}. Extending this line of inquiry, Rahman et al. further investigated non-reproducible bug reports and identified 11 key factors contributing to non-reproducibility, including missing information, ambiguous specifications, and third-party defects~\cite{rahman2020some}. Huang et al. studied invalid bug reports in deep learning compilers, revealing that such reports primarily stem from incorrect environment configuration, incorrect usage, and misunderstanding of compiler features or limitations~\cite{huang2025false}.

Regarding issues labeled as wontfix, Wang et al. analyzed 600 wontfix reports from three open-source projects on Bugzilla and identified 12 categories of reasons, including unsupported features, non-real defects, and lack of response~\cite{wang2020my}. Extending this work to GitHub, Panichella et al. constructed a dual taxonomy of opening and closing motivations, revealing that the most common reasons for wontfix decisions are feature requests deemed unnecessary and erroneous bug reports~\cite{panichella2021won}. 

These studies establish our foundational understanding of the causes of invalid issue reports and provide the basis for subsequent automated approaches. However, they primarily focus on two categories of invalid reports: bug-type invalid reports and issues labeled as wontfix, leaving the causes of invalidity for other issue types underexplored. To address this gap, we conducted a broader analysis covering invalid reports of all types to pursue a more comprehensive taxonomy of causes of invalidity.

\subsection{Machine Learning-based Classification Methods}

To automate the identification of invalid reports, researchers have explored various machine learning approaches. Zanetti et al. built classifiers based on collaboration networks, extracting features such as centrality measures from the network structure~\cite{zanetti2013categorizing}. Fan et al. expanded the feature set to 33 dimensions across five categories, i.e., reporter experience, collaboration network, completeness, readability, and text, significantly improving classification performance using random forests~\cite{fan2018chaff}. Panichella et al. employed traditional machine learning classifiers on textual features, achieving F-measures up to 0.93 and demonstrating that effective prediction is possible using only titles and descriptions~\cite{panichella2021won}.

With the advent of deep learning, He et al. proposed convolutional neural networks that classify issues based solely on titles and descriptions, eliminating the need for manual feature engineering~\cite{he2020deep}. Meng and Visser further leveraged transformer-based models such as BERT for validity classification, employing explanation methods such as LIME and SHAP to reveal key features influencing classification decisions~\cite{meng2023bug}. Evaluations of various machine learning techniques in industrial settings have shown that simple logistic regression models can meet practical needs in terms of both accuracy and interpretability~\cite{laiq2024industrial}.

However, these approaches face two limitations. First, they only output binary results without providing natural language rationales. Second, they heavily rely on high-quality training data and struggle to generalize across different projects~\cite{zimmermann2009cross, hosseini2017systematic}. These limitations call for a more explainable and generalizable approach.

\subsection{LLM-based Classification Methods}

The emergence of LLMs offers new possibilities for addressing the aforementioned limitations. Unlike traditional machine learning methods, LLMs can not only output classification results but also generate natural language rationales. Recent studies have explored applying LLMs to invalid bug report classification. Huang et al. systematically evaluated four prompting strategies and found that few-shot prompting achieves 73\% accuracy while generating explanations deemed accurate in 85.2\% of cases, demonstrating the potential of LLMs for both classification and explanation~\cite{huang2025false}. Concurrently, Dinç and Tüzün conducted a comprehensive comparison of classical ML, fine-tuned transformers, and LLMs on bug report classification~\cite{dincc2025judge}. Their fine-tuned RoBERTa achieved the highest F1, outperforming few-shot LLMs and their Judge LLM.

However, these approaches remain at the level of static text analysis and lack the capability for validity assessment through execution. Moreover, LLMs are limited by their training cutoff—information updated after training, such as project documentation and source code changes, cannot be covered by built-in model knowledge alone~\cite{vu2024freshllms}.

Meanwhile, LLM-driven agents have demonstrated emerging potential in software engineering tasks, such as program repair~\cite{yang2024swe}, code generation~\cite{zhang2024pair}, and test automation~\cite{taeb2024axnav}. Torun et al. further envisioned an AI-powered bug tracking framework where LLM agents handle tasks, including validity check~\cite{torun2025past}. These advances inspire us to explore the application of agents in issue report validity assessment.

\section{Dataset Construction} \label{dataset}
We describe how we collected and preprocessed the dataset to answer RQ1 and RQ2.

\textbf{Study Objects}: We selected three popular open-source Python projects hosted on GitHub as our data sources: pandas~\cite{pandas}, scikit-learn~\cite{scikit-learn}, and mypy~\cite{mypy}. The rationale for this selection is threefold. First, these projects have a significant influence on the software engineering community~\cite{malone2020doing,khan2021empirical}, featuring active issue tracking systems that provide rich data with both valid and invalid samples. Second, they follow established issue management practices, with maintainers clearly documenting final resolutions (e.g., closing reasons and associated pull requests), providing valuable practical evidence for establishing reliable ground truth. Third, these projects serve different technical domains: pandas for data analysis, scikit-learn for machine learning, and mypy for static type checking. This allows us to examine whether invalid issue patterns vary across software domains and enhance the generalizability of our results.

\textbf{Data Collection}: We collected issue reports from the GitHub repositories of the three selected projects using the GitHub REST API~\cite{api}, covering reports that were created since August 8, 2025 and closed by March 8, 2026. This window was chosen to ensure a sufficiently large and recent sample of issue reports for analysis. For each issue report, we extracted its ID, reporter, title, body, creation time, labels, comments, edit history for title and body, and associated pull requests (if they exist).

\textbf{Data Filtering}: To ensure the quality and representativeness of the experimental data, we applied the following filtering criteria:
\begin{enumerate}
    \item \textbf{Exclude script-generated issue reports}: We excluded reports generated by official project scripts (e.g., CI failure notifications), because these are set by the maintenance team. Only scikit-learn had such a script account (scikit-learn-bot), which submitted 45 reports.
    
    \item \textbf{Exclude maintainer-created issue reports}: Issue reports submitted by project maintainers with the ``member'' or ``collaborator'' role in the repository are typically used for tracking project maintenance activities, such as release notes and resolution discussions. There were 49 such reports in pandas, 51 in scikit-learn, and 32 in mypy.
    
    \item \textbf{Exclude issue reports without feedback}: For issue reports with neither comments nor associated pull requests, we cannot reliably determine their validity status. There were 8 such reports in pandas, 8 in scikit-learn, and 4 in mypy.
    
    \item \textbf{Exclude duplicate issue reports}: Issue reports marked as ``duplicate'' are excluded. This study primarily focuses on examining the issue content to assess its validity. There were 23 such reports in pandas, 9 in scikit-learn, and 36 in mypy.
\end{enumerate}

\textbf{Data Preprocessing}: During the lifecycle of an issue, the title or body may be edited. To ensure that our experiments evaluate the user-submitted content rather than maintainer-modified versions, we used the collected edit history to reconstruct the version before any maintainer intervention. This preprocessing avoids potential data leakage issues and ensures fairness in evaluation~\cite{tu2018careful}.

\textbf{Dataset Annotation.} After data filtering and preprocessing, we obtained 520 issue reports. Two researchers (the first and third authors), each with four years of software development experience and two years of open-source contribution experience, independently annotated the validity of each report. Based on manual examination of comments and relevant pull requests, each report was labeled as \textit{Valid} if the core claim of the report was acknowledged by community maintainers, and as \textit{Invalid} otherwise. The two annotators achieved a Cohen's kappa of 0.85, indicating substantial agreement~\cite{cohen1960coefficient,landis1977measurement}. Disagreements were resolved through discussion and, when necessary, arbitration by the second author, who has over ten years of research experience in open-source software maintenance. Table~\ref{tab:dataset-overview} provides an overview of the constructed dataset.

\begin{table}[ht]
\centering
\caption{Dataset Overview}
\label{tab:dataset-overview}
\begin{tabular}{l c c c c}
\toprule
\textbf{Project} & \textbf{Total} & \textbf{Valid} & \textbf{Invalid} & \textbf{Invalid Ratio} \\
\midrule
pandas          & 209 & 117 & 92 & 44.0\% \\
scikit-learn    & 118 & 56 & 62 & 52.5\% \\
mypy            & 193 & 90 & 103 & 53.4\% \\
\midrule
\textbf{Total}  & \textbf{520} & \textbf{263} & \textbf{257} & \textbf{49.4\%} \\
\bottomrule
\end{tabular}
\end{table}

\section{RQ1: What are the root causes of invalid issue reports across different issue types?}
\label{RQ1}

\subsection{Method}

\subsubsection{Issue Type Annotation}
To investigate whether the causes of invalid reports are correlated with issue types, we first classified the type of each issue. We observed that issue labeling practices vary across projects, and some reports lack labels altogether, making it unreliable to rely solely on existing labels for type classification. Therefore, we followed a unified classification schema based on the commonly used issue types in open-source software development, which typically include bug, enhancement, and documentation~\cite{bissyande2013got,cabot2015exploring}. The first and third authors categorized each report into one of the above types by considering the report's existing labels (when available), title, body, and comments. During this process, we observed that several issue reports fell outside these three types and further identified three additional types: CI, installation, and spam. We define the three new issue types as follows: 
\begin{itemize}
    \item CI: issue reports are related to continuous integration systems, such as pipeline failures or CI configuration defects.
    \item Installation: issue reports are related to the project installation process, such as missing pre-built packages or source code build failures.
    \item Spam: issue reports that contain no substantive content related to the project and thus cannot be meaningfully classified into any specific issue type. For example, a report may consist solely of a title like ``bug'' with no description, or contain only auto-generated template text without any user-provided information.
\end{itemize} 
We therefore incorporated these three types into our classification schema, resulting in a final set of six issue types. The two annotators achieved a Cohen's kappa of 0.88, indicating high consistency~\cite{cohen1960coefficient,landis1977measurement}. Disagreements were resolved through discussion, with the second author arbitrating when necessary.

\subsubsection{Identification of Causes of Invalidity} 
\label{thematic-analysis}
To identify the causes of invalidity, we conducted a thematic analysis on the 257 invalid issue reports, analyzing their titles, bodies, and subsequent comments. Thematic analysis is a qualitative research method used to identify, analyze, and report patterns (themes) within data~\cite{cruzes2011recommended,castleberry2018thematic}. We describe the detailed process of how we applied this approach below:

\textbf{(1) Familiarizing with the data and generating initial codes}: The two researchers independently read through all 257 invalid issue reports, including titles, bodies, and comments, to establish a holistic understanding of the data. Through a comprehensive review of the collected data, we extracted relevant phrases from each issue and transformed them into concise labels as initial codes. 
    
\textbf{(2) Searching for themes}: Within our established set of codes, we frequently observed overlapping or converging codes that addressed similar aspects. For instance, we found that codes such as ``does not align with the primary goal'' and ``better suited for a third-party library'' both suggest that the request falls outside the project's scope. We therefore merged them into an initial theme ``Out of scope''. Accordingly, we meticulously categorized these codes into groups, assigning each group a theme that summarizes their combined meaning. These themes were designed to capture the core significance of the codes within each group.

\textbf{(3) Reviewing themes}: To refine the relationships among themes, we examined each candidate theme for internal consistency and external heterogeneity. We combined related themes as sub-themes into comprehensive themes, merging, splitting, or reorganizing themes with overlapping meanings or unclear hierarchical relationships. 
    
\textbf{(4) Defining and naming themes}: When the analysis of 20 consecutive new issue reports no longer generated new codes or themes, we recognized that theoretical saturation had been reached and ceased theme organization activities~\cite{guest2006many,wutich2024sample}. In this step, we defined and named the final list of themes by confirming their exact meanings and ensuring they were named understandably and accurately.

To mitigate the influence of individual subjectivity on the analysis results, the two researchers adopted an approach involving independent coding followed by cross-validation. In cases of coding disagreements, the two researchers first attempted to reach consensus through discussion. If consensus could not be achieved after thorough discussion, the second author, who is familiar with thematic analysis and the research objectives, was invited to make the final judgment. We conducted six rounds of face-to-face discussion meetings, totaling approximately eight hours. 

\subsection{Results}
Through thematic analysis, we identified 13 causes of invalidity. Table~\ref{tab:cause-explanations} presents these causes along with their detailed explanations and their distribution across the three studied projects. We can see that, despite differences in domain and scale among pandas, scikit-learn, and mypy, the same causes of invalidity appear across all three projects, indicating their potential for cross-project generalization. 

\begin{table*}[ht] 
\caption{Root Cause Taxonomy and Project Distribution of Invalid Issue Reports}
\label{tab:cause-explanations}
\begin{tabularx}{\textwidth}{l X c c c}
\toprule
\textbf{Cause} & \textbf{Explanation} & \textbf{pd} & \textbf{sk} & \textbf{mp} \\
\midrule
User misunderstanding & The issue stems from user misunderstanding of the project's expected behavior, such as intentional code behavior, correct documentation, or reasonable project configuration. & 26 & 15 & 56 \\
Unable reproduction & The issue cannot be reproduced based on the information provided in the report. & 11 & 5 & 10 \\
Misplaced report & The issue stems from a third-party project, not in the project itself. & 12 & 4 & 8 \\
Out of date & The issue has been fixed in subsequent versions. & 4 & 1 & 8 \\
Inherent limitation & Due to inherent constraints in the project's technology stack, the issue cannot be resolved. & 1 & 2 & 5 \\
Existing alternative & The request can already be achieved through other means. & 13 & 9 & 2 \\
Out of scope & The request does not align with the project or is better implemented as an external project. & 10 & 9 & 1 \\
Backward incompatibility & The requested change will cause backward incompatibility. & 2 & 2 & 2 \\
Poor ROI & The request is too niche or too complex to implement with limited utility. & 1 & 3 & 4 \\
Adequate documentation & Current documentation already clearly covers the relevant content. & 2 & 4 & 1 \\
Unsupported version & The issue stems from attempting to install officially unsupported versions. & 1 & 1 & 1 \\
Environment misconfiguration & The issue stems from user-specific permissions, dependency conflicts, or incorrect environment settings. & 1 & 2 & 3 \\
Non-informative & The issue is extremely incomplete, lacking sufficient information to be understood. & 8 & 5 & 2 \\
\bottomrule
\multicolumn{5}{l}{\textit{Note:} pd = pandas, sk = scikit-learn, mp = mypy.} \\
\end{tabularx}
\end{table*}

A prior study~\cite{panichella2021won} has identified 26 closing reasons for wontfix issues. When compared with these reasons, most have counterparts in our framework, such as user misunderstanding, out of scope, and existing alternative. However, three causes we repeatedly observed are not covered by their taxonomy: adequate documentation, misplaced report, and non-informative. These additions stem from our dataset covering a broader range of issue types, such as documentation and spam. 

Further, we analyzed the relationship between the identified causes of invalidity and issue types. The correspondence between them is shown in Figure \ref{fig:cause-type}. We can observe that the distribution of causes of invalidity exhibits distinct patterns across different issue types:

\begin{figure}[ht]
\centering
\includegraphics[width=0.5\textwidth]{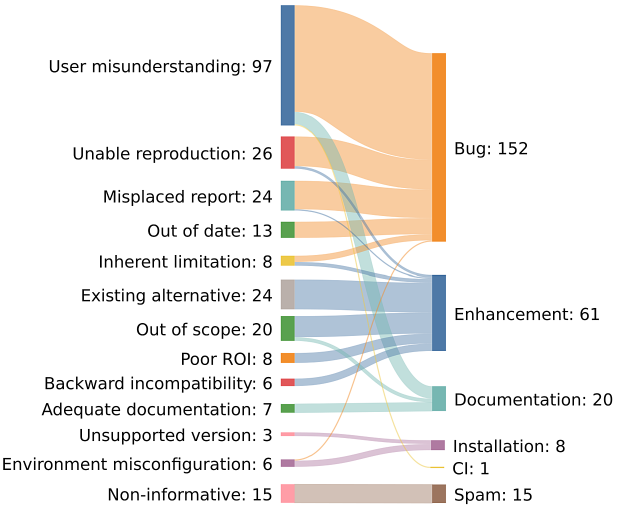}
\caption{Correspondence Between Causes of Invalidity and Issue Types}
\label{fig:cause-type}
\end{figure}

\textbf{Bug} type has multiple causes of invalidity, dominated by ``User misunderstanding'', followed by ``Unable reproduction'', ``Misplaced report'', ``Out of date'', and ``Inherent limitation''. This indicates that bug reports are most often deemed as invalid due to the reporter's misunderstanding of the expected behavior, or the reproduction example is missing or incorrect, rather than an actual defect in the project itself.

\textbf{Enhancement} type has the most diverse set of causes of invalidity, with the most common being ``Existing alternative'' and ``Out of scope'', followed by ``Backward incompatibility'' and ``Poor return on investment (ROI)''. This suggests that enhancement reports are deemed invalid primarily because the proposed functionality is inconsistent with the project's core goals, already achievable through other means, or lacks implementation value from a maintenance cost perspective.

\textbf{Documentation} type shows a relatively concentrated pattern, mainly consisting of three categories: ``User misunderstanding'', ``Adequate documentation'', and ``Out of scope''. This indicates that such invalid reports primarily stem from reporters' misunderstanding of the documentation, overlooking existing documentation, or misalignment with the scope of the documentation.

\textbf{Installation} type includes two causes of invalidity: ``Unsupported version'' and ``Environment misconfiguration''. Such reports are deemed invalid because users either use officially unsupported installation packages or have specific configuration issues in their local environment.

\textbf{CI} type has a single cause of invalidity, namely ``User misunderstanding'', with this report stemming from the user's misunderstanding of the expected behavior of the CI system.

\textbf{Spam} type has a single cause of invalidity, namely ``Non-informative'', with such reports being extremely incomplete and exhibiting clear characteristics. 

This finding reveals a strong correlation between issue types and causes of invalidity. Specifically, ``User misunderstanding'' and ``Unable reproduction'', which are common in bug reports, rarely appear in enhancement reports; conversely, ``Out of scope'' and ``Existing alternative'', which are common in enhancement reports, do not appear in bug reports. 

\vspace{0.5em}
\noindent
\fcolorbox{black}{gray!15}{
\begin{minipage}{0.96\columnwidth}
\textbf{Summary for RQ1:} We identified 13 causes of invalidity from 257 invalid reports, clarifying the specific reasons that lead issue reports to be deemed invalid. The distribution of causes of invalidity varies significantly across issue types, revealing a strong correlation between issue types and these causes.
\end{minipage}
}

\section{RQ2: What is the Performance of \appname Compared to the Baselines?}
\label{RQ2}
Inspired by the findings of RQ1, we thoroughly considered the causes of invalidity and types of issue reports to design an automated approach of issue report validity assessment, noted as \appname.

\subsection{Approach Design}

\begin{figure}
    \centering
    \includegraphics[width=0.5\textwidth]{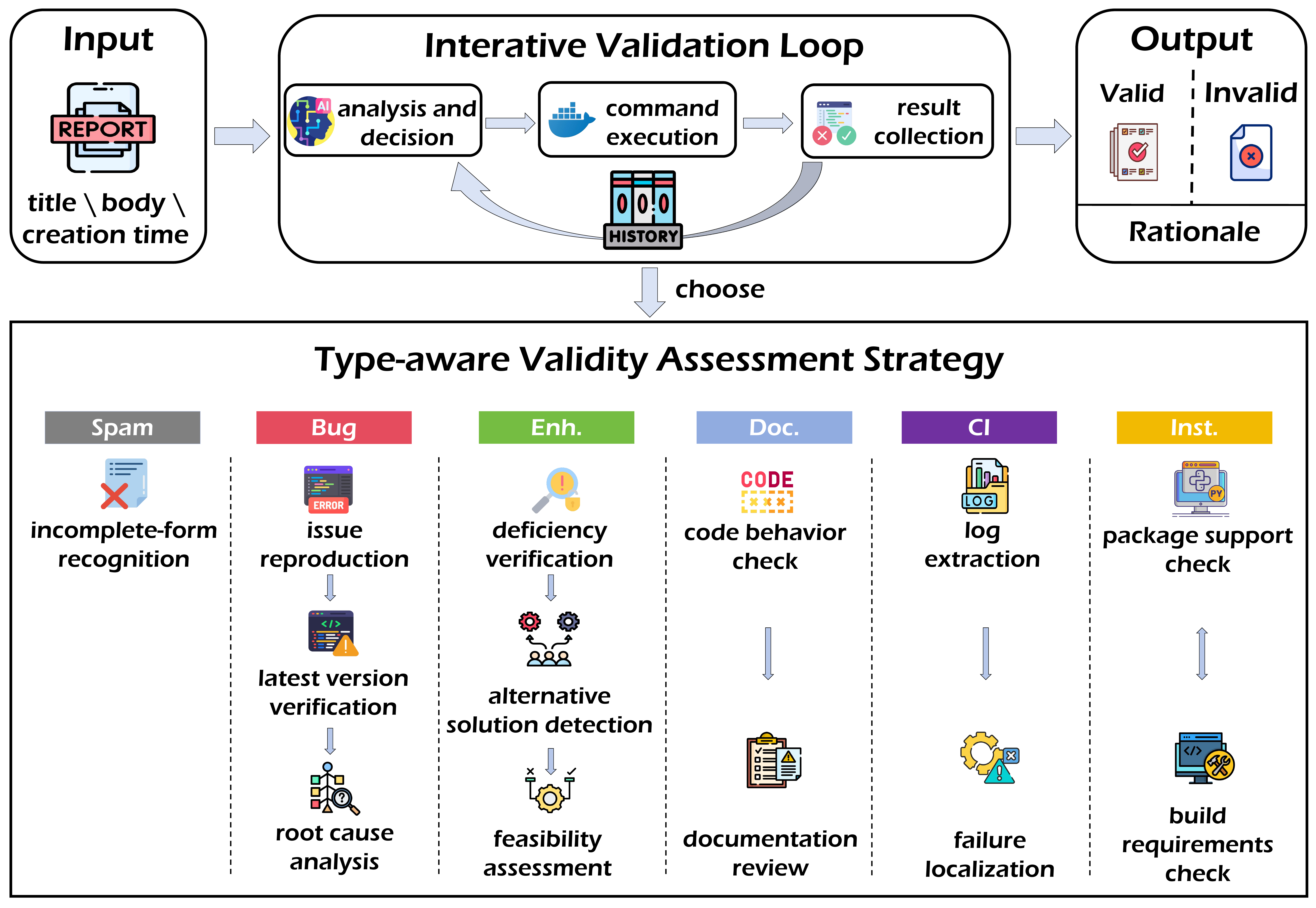}
    \caption{Overall Architecture of \appname}
    \label{fig:architecture}
\end{figure}

Figure~\ref{fig:architecture} shows the overall architecture of \appname. Our approach takes issue report information, i.e., title, body, and creation time, as input, validates it through an interactive loop with a type-aware validity assessment strategy, and ultimately outputs the validity judgment of the issue report along with its rationale. The detailed design of \appname is described as follows.

\subsubsection{Interactive Validation Loop}

Based on the 13 causes of invalidity identified in RQ1, we found that the determination of most causes requires information beyond the issue report itself and varies among issue types. For example, ``Unable Reproduction'' requires actual code execution to confirm, while ``User Misunderstanding'' requires consulting official documentation to assess whether the expected behavior is reasonable. These determinations collectively point to a core requirement: acquiring diverse information beyond issue reports. LLM-based agents naturally possess this capability~\cite{xi2025rise}. Based on these observations, we designed \appname as an LLM-based agent with an interactive validation loop as its core operating mechanism, empowering it with the ability to execute commands.

\appname operates through a cycle of ``analysis and decision $\rightarrow$ command execution $\rightarrow$ result collection $\rightarrow$ continue analysis and decision''~\cite{yang2023intercode}, with all interaction history accumulated as context for subsequent analysis. \appname follows a structured output format: (1) an analysis of the current state; (2) a status indicator (CONTINUE or END); and (3) the corresponding content. If the status is CONTINUE, the agent outputs executable bash commands; if END, it outputs the final validity judgment along with its rationale. This structured output enables the agent to iteratively gather evidence and autonomously terminate the loop when sufficient information has been collected.

To prevent infinite loops during validation, we set a maximum number of interaction rounds as a safety upper bound. In preliminary experiments on 105 randomly sampled reports (35 from each project), over 95\% of validations were completed within 9 interaction rounds. Accordingly, we set the upper bound to \textbf{9}. When \appname determines that sufficient evidence has been gathered to reach a conclusion, it autonomously ends the loop; if the maximum number of interaction rounds is reached, it forcibly terminates the loop and renders a judgment based on the evidence collected so far. In our main experiments, the average number of interaction rounds for ChatGPT, Grok, and Qwen is 2.8, 4.1, and 4.4, respectively, with the corresponding 95th percentiles at 6, 8, and 8. This indicates that the 9-round upper bound is sufficient for the vast majority of validations.

To ensure isolation and reproducibility of the validation environment, each validation session runs in an independent Docker container~\cite{merkel2014docker}. Our Docker image contains the full source code of the target project and is equipped with Git for version switching, Conda for environment management, pip for dependency installation, and grep for code and documentation retrieval. Since not all issue reports specify the affected code version, we use the GitHub REST API to retrieve the latest main-branch commit preceding each report's creation time as a fallback version reference. During validation, \appname checks out the reported version when available; otherwise, it uses the retrieved commit for creating a Conda environment and installing the required dependencies.

\subsubsection{Type-Aware Validity Assessment Strategy}

Furthermore, the findings of RQ1 motivate us to consider that the information dimensions required for validity assessment vary across issue types. An ideal validity assessment strategy should adopt differentiated approaches tailored to each issue type, rather than a one-size-fits-all approach. Based on this, we further analyzed the determination requirements of each cause of invalidity under each type and designed targeted validation steps for each issue type. \appname first classifies each issue report into one of six types based on its title and body to select a validity assessment strategy. In case \appname considers the report to lie between multiple types, it will draw on multiple strategies as needed until sufficient evidence is gathered. The detailed design is as follows.

For \textit{bug} type, reproducibility is the foundation of validity~\cite{zimmermann2010makes, jin2012bugredux} and ranks among the top causes of invalidity as shown in Figure \ref{fig:cause-type}. Therefore, the validity assessment process first attempts to reproduce the issue based on the information provided in the report. If reproduction fails, the issue is judged as ``Unable reproduction''. Otherwise, \appname further verifies whether the issue still exists in the latest version: if it has been fixed, the issue is judged as ``Out of date''; if it persists, the agent conducts an in-depth root cause analysis, combining code behavior and documentation to determine whether the issue stems from user misunderstanding of expected behavior, belongs to a third-party project, or is an inherent limitation of the project itself.

For \textit{enhancement} type, validity assessment centers on assessing the reasonableness and feasibility of the request. Since enhancement requests are typically motivated by a perceived deficiency, the process first verifies whether the claimed deficiency actually exists. If not, the request lacks a basis and requires no further evaluation. Next, \appname checks whether existing features already provide an alternative solution, as such redundancy would render the request unnecessary. Finally, it evaluates the request's alignment with the project's design principles and maintenance considerations, identifying potential issues such as being out of scope, having poor ROI, or introducing backward incompatibility.

For \textit{documentation} type, the core lies in verifying the correctness and completeness of the documentation. If the documentation involves functional behavior (such as API usage examples or code comments), \appname executes the relevant code to verify the actual behavior. Otherwise, it directly retrieves the relevant documentation. Through comparison, if the documentation is correct but the user's expectation deviates, it is judged as ``User misunderstanding''. If the documentation accurately covers the relevant content, the issue is judged as ``Adequate documentation''. If the documentation contains errors or omissions, further judgment is required to determine whether it falls outside the scope of documentation maintenance.

For \textit{CI} type, the core lies in identifying the cause of failure. If the report contains a CI log link, \appname will first extract the content to obtain detailed error information; if not, it relies on the information provided in the report. Then, it analyzes the specific failure cause and consults the project's CI configuration to comprehensively determine whether the failure stems from user misunderstanding or issues with the CI system itself.

For \textit{installation} type, the core lies in distinguishing whether the source of the issue is the project itself or the user. If the report is related to pre-built packages, \appname checks whether the package used in the report falls within the project's official support scope. If the report is related to the source build, it consults the project's installation guide to determine whether the user has met the installation requirements, and checks for user-specific issues, such as permissions or dependency conflicts.

For \textit{spam} type, reports are characterized by extremely incomplete content and are easy to identify. Therefore, \appname only assesses the report content and directly judges it as invalid upon recognition, without further processing. 

\subsection{Experimental Setup}

\subsubsection{Baselines}

We compare \appname against three representative state-of-the-art baselines covering both prompting-based and learning-based approaches. For prompting-based methods, Huang et al. systematically evaluated four strategies, i.e., zero-shot, few-shot, chain-of-thought, and retrieval-augmented generation. Their results showed that enhanced zero-shot achieved the highest recall and F1-score, while few-shot achieved the highest precision and accuracy. Chain-of-thought limited LLM adaptability due to rigid stepwise reasoning, and retrieval-augmented generation suffered from noisy and version-misaligned documentation retrieval~\cite{huang2025false}. We therefore include only enhanced zero-shot and few-shot as baselines. As a representative learning-based baseline, we include the fine-tuned RoBERTa model proposed by Dinç and Tüzün~\cite{dincc2025judge}, which reported the best performance among traditional classifiers. These three baselines provide comprehensive coverage of the strongest existing approaches for issue report validity assessment.

The enhanced zero-shot and few-shot baselines adopt the same validity criteria as \appname. For few-shot prompting, we follow the dynamic selection strategy of Huang et al.~\cite{huang2025false}. Specifically, for each project, we constructed a balanced seed pool of 26 examples selected from the preliminary experiment set: 13 invalid cases, each representing one of the 13 causes of invalidity identified in RQ1, and 13 valid cases covering issue types. These 78 examples were excluded from the final evaluation set, leaving 442 reports for all testing. For each test report, we randomly sampled one invalid and one valid example from the seed pool as in-context examples. For the fine-tuned RoBERTa baseline, we follow the fine-tuning setup of Dinç and Tüzün~\cite{dincc2025judge}. We collected 2,000 issue reports per project created before August 8, 2025, and labeled each as valid if it was resolved with a merged pull request, and as invalid otherwise. We then used this data to fine-tune RoBERTa for each project. The fine-tuned model was then evaluated on the same test set used for \appname and the prompting-based baselines.

\subsubsection{LLM Models}
We implement both \appname and the prompting-based baselines using three mainstream large language models: \textbf{ChatGPT}, \textbf{Grok}, and \textbf{Qwen}. These models represent a diverse set of state-of-the-art LLMs with strong performance on natural language understanding and code-related tasks. ChatGPT (gpt-5) represents leading closed-source commercial models~\cite{singh2025openai}, Grok (grok-4) represents models optimized for reasoning capabilities~\cite{xai2025grok4}, and Qwen (qwen3-235b-a22b-instruct-2507) represents powerful open-source alternatives~\cite{yang2025qwen3}. All model versions were released prior to our data collection period (August 8, 2025), ensuring that the models have not been exposed to the evaluation data during training. For all LLM calls in both methods, we set the temperature to \textbf{0} to maximize reproducibility by having the model select the most likely token at each step. All other hyperparameters were kept at their API default values~\cite{renze2024effect}.

\subsubsection{Evaluation Metrics}

Since our primary goal is to identify invalid issue reports, we treat \textit{invalid as positive} and \textit{valid as negative}. For each method, we report accuracy, precision, recall, and F1-score.

\subsection{Results}

We evaluate \appname from two perspectives: (1) quantitatively assess its classification performance, including overall comparison with the baselines, performance on each issue type, ablation studies to assess the contribution of each component, and cost-runtime analysis; (2) to further examine whether the type-aware validity assessment strategy proved effective, we conducted a qualitative analysis from three aspects: consistency between model-generated rationales and human-annotated causes of invalidity for correctly identified invalid reports, rationale characteristics for correctly identified valid reports, and failure patterns of misclassified cases.

\subsubsection{Quantitative Evaluation}

Table~\ref{tab:performance-comparison} shows the overall performance comparison between \appname and the three baseline methods. The experimental results demonstrate that \appname significantly outperforms all baselines across all evaluation metrics, and this advantage remains consistent across all three LLMs. Taking ChatGPT as an example, \appname achieves an accuracy of 96.2\% and an F1-score of 96.0\%, compared to enhanced zero-shot (78.5\% accuracy, 74.5\% F1) and few-shot (78.3\% accuracy, 76.4\% F1). On Grok, \appname achieves an F1 of 86.6\%, versus 55.5\% and 57.8\% for enhanced zero-shot and few-shot, respectively. On Qwen, \appname achieves an F1 of 83.2\%, while enhanced zero-shot and few-shot achieve 56.4\% and 64.4\%, respectively. The fine-tuned RoBERTa achieves an F1 of 65.9\%, lower than \appname on all three models. These results confirm that our agent-based approach effectively overcomes the limitations of static text analysis, providing a more reliable solution for issue report validity assessment.

\begin{table}[ht]
\centering
\caption{Performance Comparison of \appname and Baseline Methods}
\label{tab:performance-comparison}
\begin{tabular}{l l l c c c c}
\toprule
\textbf{Model} & \textbf{Method} & \textbf{Acc.} & \textbf{Prec.} & \textbf{Rec.} & \textbf{F1} \\
\midrule
\multirow{3}{*}{ChatGPT} & \appname & \textbf{96.2\%} & \textbf{98.6\%} & \textbf{93.6\%} & \textbf{96.0\%} \\
 & Enhanced zero-shot & 78.5\% & 89.7\% & 63.8\% & 74.5\% \\
 & Few-shot & 78.3\% & 84.4\% & 69.7\% & 76.4\% \\
\cmidrule(lr){1-6}
\multirow{3}{*}{Grok} & \appname & \textbf{88.0\%} & \textbf{96.6\%} & \textbf{78.4\%} & \textbf{86.6\%} \\
 & Enhanced zero-shot & 68.8\% & 93.5\% & 39.4\% & 55.5\% \\
 & Few-shot & 68.3\% & 84.2\% & 44.0\% & 57.8\% \\
\cmidrule(lr){1-6}
\multirow{3}{*}{Qwen} & \appname & \textbf{85.3\%} & \textbf{95.3\%} & \textbf{73.9\%} & \textbf{83.2\%} \\
 & Enhanced zero-shot & 68.6\% & 89.1\% & 41.3\% & 56.4\% \\
 & Few-shot & 71.7\% & 85.0\% & 51.8\% & 64.4\% \\
 \midrule
RoBERTa & \multicolumn{1}{c}{—} & 65.6\% & 66.2\% & 65.6\% & 65.9\% \\
\bottomrule
\end{tabular}
\end{table}

We further investigate the performance of \appname across different issue types on three models. Table~\ref{tab:performance-by-type} presents the accuracy and F1-scores for each issue type on ChatGPT, Grok, and Qwen. The results show that \appname achieves robust performance on the three most common types, i.e., bug, enhancement, and documentation, across all models. For the remaining types (CI, installation, and spam), despite the limited sample sizes, the results remain promising.

\begin{table}[ht]
\centering
\caption{Performance by Issue Types Across Models}
\label{tab:performance-by-type}
\begin{tabular}{l c c c c c c c}
\toprule
\multirow{2}{*}{Type} & \multirow{2}{*}{N} & \multicolumn{2}{c}{ChatGPT} & \multicolumn{2}{c}{Grok} & \multicolumn{2}{c}{Qwen} \\
\cmidrule(lr){3-4} \cmidrule(lr){5-6} \cmidrule(lr){7-8}
 & & Acc. & F1 & Acc. & F1 & Acc. & F1 \\
\midrule
Bug & 296 & 96.6\% & 96.9\% & 86.5\% & 88.5\% & 87.2\% & 89.1\% \\
Enh. & 65 & 93.8\% & 89.5\% & 90.8\% & 85.0\% & 81.5\% & 73.9\% \\
Doc. & 58 & 96.6\% & 97.7\% & 89.7\% & 93.0\% & 77.6\% & 85.4\% \\
CI & 6 & 100\% & 100\% & 83.3\% & 90.9\% & 83.3\% & 90.9\% \\
Inst. & 5 & 80.0\% & 66.7\% & 100\% & 100\% & 80.0\% & 66.7\% \\
Spam & 12 & 100\% & 100\% & 100\% & 100\% & 100\% & 100\% \\
\bottomrule
\end{tabular}
\end{table}

\textbf{Ablation Study.} In RQ1, we have two findings: a taxonomy of 13 causes of invalidity and a strong correlation between these causes and issue types. Based on these, \appname incorporates two components: the taxonomy as explicit judgment criteria and a type-aware validity assessment strategy designed on the observation that different issue types exhibit distinct patterns of invalidity. 
To assess the contribution of each component, we designed two ablation variants: (1) \textbf{w/o Strategy}, which removes the type-aware validity assessment strategy while retaining the taxonomy; and (2) \textbf{w/o Strategy \& Taxonomy}, which removes both components, retaining only the agent's execution capability with a generic definition of validity.

\begin{table}[ht]
\centering
\caption{Ablation Study Results}
\label{tab:ablation-results}
\setlength{\tabcolsep}{4pt}
\begin{tabular}{l l c c c c}
\toprule
\textbf{Model} & \textbf{Variant} & \textbf{Acc.} & \textbf{Prec.} & \textbf{Rec.} & \textbf{F1} \\
\midrule
\multirow{2}{*}{ChatGPT}
 & w/o Strategy & 82.5\% & 84.4\% & 79.4\% & 81.8\% \\
 & w/o Strategy \& Taxonomy & 76.5\% & 95.2\% & 55.0\% & 69.8\% \\
\midrule
\multirow{2}{*}{Grok}
 & w/o Strategy & 73.7\% & 82.3\% & 59.6\% & 69.1\% \\
 & w/o Strategy \& Taxonomy & 67.4\% & 99.9\% & 33.9\% & 47.8\% \\
\midrule
\multirow{2}{*}{Qwen}
 & w/o Strategy & 74.8\% & 85.9\% & 58.7\% & 69.8\% \\
 & w/o Strategy \& Taxonomy & 70.4\% & 93.1\% & 43.1\% & 58.9\% \\
\bottomrule
\end{tabular}
\end{table}

Table~\ref{tab:ablation-results} shows that removing only the type-aware validity assessment strategy leads to an average F1-score decline of approximately 14.2 percentage points across the three LLMs. Removing both components results in a larger decline of approximately 22.4 percentage points. This variant achieves high precision but low recall, suggesting that without the taxonomy, the agent lacks clear guidance on what constitutes an invalid report. These results confirm that both the type-aware strategy and the taxonomy contribute to the effectiveness of \appname.

\textbf{Cost and Runtime Analysis.} We further compared the average API call cost and runtime of each method, as shown in Table~\ref{tab:cost-runtime}. In terms of cost, \appname incurs higher API fees than enhanced zero-shot and few-shot. However, considering the substantial performance gains and the high effort of manual review, the additional cost seems acceptable in practice. Regarding runtime, \appname takes significantly longer than enhanced zero-shot and few-shot, primarily because it requires multiple rounds of interaction to configure the virtual environment, execute code, and retrieve information. Further analysis reveals that, under the ChatGPT configuration, environment reconstruction takes approximately 70 seconds on average, accounting for about 59\% of \appname's total runtime. If issue reporters have already configured the environment locally, \appname can reuse it directly, reducing the wait to roughly 50 seconds, which is more acceptable for a real-time submission scenario. For maintainers, checking the validity of issue reports is not a latency-critical scenario. Moreover, since validation tasks are independent, multiple agents can be configured to process concurrent submissions in parallel. Notably, both the cost and runtime of the two ablation variants are close to those of the full version, indicating that the taxonomy and type-aware strategy introduce little additional overhead.

\begin{table*}[ht]
\centering
\caption{Average Cost (\$) and Runtime (s) per Issue Report}
\label{tab:cost-runtime}
\begin{tabular}{l c c c c c c}
\toprule
\multirow{2}{*}{Method} & \multicolumn{2}{c}{ChatGPT} & \multicolumn{2}{c}{Grok} & \multicolumn{2}{c}{Qwen} \\
\cmidrule(lr){2-3} \cmidrule(lr){4-5} \cmidrule(lr){6-7}
 & cost (\$) & runtime (s) & cost (\$) & runtime (s) & cost (\$) & runtime (s) \\
\midrule
\appname & 0.045 & 119.4 & 0.042 & 96.3 & 0.007 & 106.4 \\
Enhanced zero-shot & 0.019 & 23.3 & 0.005 & 5.8 & 0.001 & 4.6 \\
Few-shot & 0.021 & 23.6 & 0.010 & 6.0 & 0.001 & 5.0 \\
w/o Strategy & 0.044 & 110.1 & 0.039 & 82.4 & 0.006 & 102.5 \\
w/o Strategy \& Taxonomy & 0.045 & 95.1 & 0.040 & 78.9 & 0.006 & 98.7 \\
\bottomrule
\end{tabular}
\end{table*}

\subsubsection{Qualitative Evaluation}

We complemented quantitative metrics with a qualitative analysis of \appname on ChatGPT (the best-performing model) from three perspectives to examine whether the type-aware validity assessment strategy proved effective. First, we examined whether \appname correctly identifies why an issue report is invalid by comparing the generated rationale with what we annotated in RQ1. Second, we analyzed the characteristics of its rationales when \appname correctly classifies an issue as valid, to investigate whether its judgments are based on findings from the validity assessment process or merely on the original report text. Following the annotation procedure described in Section~\ref{dataset}, the two annotators achieved Cohen's kappa of 0.89 and 0.83 for the rationale analyses of invalid and valid reports, respectively. Third, we performed a failure analysis using the same analysis method described in Section~\ref{thematic-analysis} to understand the limitations of \appname.

\textbf{Alignment with Annotated Causes of Invalidity.} We randomly sampled 100 reports that \appname correctly classified as invalid. We compared the rationales generated by \appname against the causes of invalidity we annotated in RQ1, categorizing the alignment into three types: (1) exact match, where the core reason in the rationale aligns with the annotated cause; (2) partial match, where the rationale includes the annotated cause but also mentions additional factors; and (3) mismatch, where the rationale identifies a different cause. Among the 100 samples, 69 (69\%) were exact matches, 28 (28\%) were partial matches, and 3 (3\%) were mismatches. This indicates that \appname's determinations of invalidity are strongly aligned with the underlying causes, rather than occurring by chance. These findings suggest that the provided rationale can guide users to adjust their issue reports. 

\textbf{Rationale for Correctly Classified Valid Reports.} Similarly, we randomly sampled 100 reports that \appname correctly classified as valid and analyzed the characteristics of its generated rationales. The results show that in 99 cases (99\%), the rationales explicitly mentioned key results from the validity assessment process that confirmed the authenticity of the issue. In 54 cases (54\%), the rationales further pinpointed the exact location of the issue, such as specific code blocks or documentation sections. These characteristics demonstrate that when judging an issue report as valid, \appname's rationales are grounded in empirical evidence, obtaining first-hand information through code execution, documentation retrieval, and other means rather than reasoning over the report text.

\textbf{Failure Analysis.} The 17 cases where \appname made incorrect validity judgments fell into five categories: key information not retrieved or misinterpreted (7 cases); failure to identify undocumented design decisions (3 cases); lack of practical engineering considerations (3 cases); reproduction failures caused by environmental discrepancies or dependency changes (2 cases); and misattribution of fix responsibility between the project and third-party dependencies (2 cases). The results indicate that most errors fundamentally stem from limitations in information retrieval and interpretation, which is the key area to improve in the future.

During the failure analysis, we identified four cases where \appname's rationale revealed factual evidence that had not been mentioned in the original discussion, such as discrepancies between documentation and actual releases or inconsistencies in code behavior. This evidence supported a validity judgment different from the maintainers' initial decision. In all cases, \appname judged the reports valid, while the maintainers had initially closed them as invalid. For each case, we extracted the key evidence from \appname's final rationale and reorganized it into a new issue report submitted to the respective GitHub repository. All four cases resulted in positive outcomes: maintainers acknowledged the feedback and ultimately merged pull requests addressing the issues. This further demonstrates that the rationales generated by \appname can provide sufficiently clear and actionable evidence to assist maintainers in their decision-making.

\vspace{0.5em}
\noindent
\fcolorbox{black}{gray!15}{
\begin{minipage}{0.96\columnwidth}
\textbf{Summary for RQ2:} 
\appname consistently outperforms all baseline methods in validity assessment effectiveness. Performance across different issue types remains stable across all LLMs. Ablation studies further confirm that both the taxonomy of causes of invalidity and the type-aware validity assessment strategy contribute to the performance gains. Qualitative analysis confirms that its rationales align closely with annotated causes of invalidity and are grounded in empirical evidence. Across four cases where our assessments of issue report validity disagreed with maintainers, we submitted appeals supported by evidence provided by \appname; all were accepted, indicating the practical utility of our approach. 
\end{minipage}
}

\section{Implications}
\label{implications}
This section discusses the implications of our study for researchers, maintainers, and reporters.

\textbf{Implications for Researchers.} 
Our study provides the first investigation of causes of invalidity for all types of issue reports, establishing a taxonomy of 13 causes of invalidity across six issue types. This taxonomy lays a theoretical foundation for future research on invalid report analysis. Researchers can build upon it to conduct more fine-grained cause analysis or to investigate the causes of invalidity in other software ecosystems. Additionally, our manually annotated dataset of 520 issue reports (including annotations of causes of invalidity for 257 invalid reports) can serve as a benchmark for evaluating automated issue report validity assessment techniques, providing empirical data to support future studies. Our failure analysis further reveals five categories of challenges encountered in our approach, such as improving information retrieval robustness and better handling of issues arising from cross-project dependencies, which is a promising research avenue.

\textbf{Implications for Maintainers.}
Our study reveals that invalid issue reports are not random noise but follow predictable patterns tied to specific issue types. This insight enables maintainers to move from reactive handling toward proactive prevention. For instance, given that ``User misunderstanding'' is the most common cause among bug reports, maintainers can improve documentation by clarifying examples or design issue templates that prompt users to confirm whether they have consulted relevant documentation. For enhancement requests, which are often invalid due to ``Existing alternative'' or ``Out of scope'', projects can establish clearer contribution boundaries and provide self-service checklists to help reporters assess alignment with project goals before submission. Furthermore, the \appname framework demonstrates the feasibility of agent-based automated validity assessment; maintainers can draw on this approach to develop assistive tools that automatically detect common causes of invalidity at submission time or provide reference judgments during issue processing, thereby reducing manual overhead.

\textbf{Implications for Issue Reporters.} 
Our study reveals that a substantial portion of invalid issue reports stems from reporters' misunderstanding of project documentation, code, or scope, as well as missing or incorrect critical information. Reporters can increase the likelihood of their issue reports being valid by first carefully consulting relevant documentation to confirm whether the observed behavior aligns with project expectations; second, understanding the project's core scope and design goals to avoid submitting out-of-scope requests; and third, providing complete and accurate information (e.g., a minimal reproduction example). These practices not only reduce maintainers' burden of handling invalid reports but also significantly improve the chances that reports will be promptly addressed and resolved. In essence, a high-quality issue report is itself a contribution to the project community, because it saves maintainers time that would otherwise be spent clarifying misunderstandings, allowing them to focus on solving genuine issues.

\section{Threats to Validity}
\label{threats}
This section discusses potential internal and external threats to the validity of our study.

\subsection{Internal Validity}

\textbf{Limited Sample Size for Certain Issue Types.} In our dataset, some issue types, particularly CI and Installation, have relatively few samples. This limited sample size may affect the comprehensiveness of our analysis of causes of invalidity for these types and could impact the reliability of experimental results for these types. Future work with larger datasets encompassing more projects is needed to further validate our findings on these less common issue types.

\textbf{Subjectivity.} The annotation of issue types and validity, thematic analysis of causes of invalidity, and qualitative evaluation all involve subjective judgment, which may introduce researcher bias. To mitigate this threat, two researchers independently performed all annotation and coding tasks, with disagreements resolved through discussion and arbitration by a third researcher. While these processes do not eliminate subjectivity entirely, they reduce the risk of individual bias.

\textbf{Randomness of LLM Outputs.} The outputs of large language models are inherently stochastic, which may affect the reproducibility of experimental results. To mitigate this effect, we set the temperature parameter to 0 in all experiments, making the model outputs as deterministic as possible~\cite{renze2024effect}. While setting the temperature to 0 does not guarantee identical outputs across multiple runs, it minimizes the impact of randomness on the reproducibility of our experimental results.

\textbf{Label Criterion Discrepancy for RoBERTa.} For the RoBERTa baseline, we followed the resolution-to-label mapping strategy of Dinç and Tüzün~\cite{dincc2025judge}. However, because GitHub has no explicit field equivalent to Bugzilla’s resolution, and manually annotating roughly 6,000 training reports is infeasible, we used merged pull requests as a proxy signal for validity. Applying the same rule to our annotated dataset yields only 11 discrepancies (about 2.1\%). This introduces small noise that underestimates RoBERTa’s performance. However, the core comparison of this paper is between LLM-based methods; RoBERTa serves only as a reference
for traditional learning approaches. Therefore, this minor label variance does not affect the validity of our main conclusions.

\textbf{Data Overlap.} The taxonomy was constructed from all 257 invalid reports, primarily based on maintainers' comments, whereas validity assessment relies on issue content. Although the two processes involve overlapping reports, they rely on different sources of information. Additionally, 105 reports were used in preliminary experiments to refine the type-aware strategy and determine the interaction round limit. After excluding 78 few-shot examples, 27 reports remained in the final evaluation set. To assess the impact of this overlap, we separately evaluated these 27 reports and the remaining 415 reports using ChatGPT, obtaining accuracy and F1-score of 92.6\% and 92.3\%, and 96.4\% and 96.2\%, respectively. These results show no apparent performance inflation on the overlapping reports.

\textbf{Interaction Round Limit.} The maximum number of interaction rounds was set to 9 based on preliminary experiments. In the main experiments, 2, 12, and 12 cases reached this limit for ChatGPT, Grok, and Qwen, respectively, of which 0, 2, and 4 were incorrectly assessed. This fixed limit may constrain the assessment of complex issues, motivating future work on adaptive termination strategies.

\subsection{External Validity}

\textbf{Generalizability Across Projects.} We selected three representative open-source Python projects from GitHub, i.e., pandas, scikit-learn, and mypy, for our study. Although these projects are influential in the software engineering community and maintain active issue tracking systems, they cannot represent all types of software projects, particularly small or personal projects, projects written in other programming languages, or industrial closed-source projects~\cite{kalliamvakou2014promises}. To examine the potential generalizability of our taxonomy beyond the selected projects, we compared it with prior studies on C\# projects~\cite{panichella2021won}, Java projects~\cite{sun2011bug}, Bugzilla projects~\cite{wang2020my}, and deep learning compiler projects~\cite{huang2025false}. We found that most of the core causes of invalidity reported in these studies could be mapped to our 13-category taxonomy (detailed comparisons are provided in our replication package). This suggests that the core causes of invalidity exhibit some commonality across different ecosystems, indicating the potential generalizability of our taxonomy. Furthermore, the evaluation of \appname is limited to our dataset, and its effectiveness on projects with different characteristics remains to be investigated. In future work, we will further expand our dataset to cover diverse programming languages, domains, and scales to validate the applicability of our taxonomy, identify potential new causes, and further evaluate the effectiveness of \appname.

\textbf{Stability of Issue Status.} Our dataset consists of issue reports created since August 8, 2025 and closed by March 8, 2026, and we annotated their validity status based on their state at the time of data collection (March 9, 2026). However, the resolution status of an issue is not always final. Some closed issues may be reopened in the future~\cite{shihab2013studying}. This means that some of our validity judgments, made based on the state at a single point in time, could potentially be overturned if issues are later reopened and addressed.

\section{Conclusion}
\label{conclusion}
This study advances the understanding and automated identification of invalid issue reports from both empirical and methodological perspectives. On the empirical side, through a thematic analysis of 257 invalid reports, we established the taxonomy of 13 causes of invalidity across six issue types, revealing the strong correlation between causes of invalidity and issue types. On the methodological side, to address the fundamental limitations of existing automated approaches, we propose \appname, an agent-based framework with a type-aware validity assessment strategy. By dynamically executing commands, \appname transforms issue validity assessment from a text-based inference paradigm to an evidence-based verification paradigm. Quantitative results across three LLMs show that \appname consistently outperforms all baselines by a substantial margin. The performance remains robust across different issue types, and ablation studies confirm the contribution of both the type-aware validity assessment strategy and the taxonomy of causes of invalidity. More importantly, qualitative analysis confirms that \appname's judgments are built upon a substantive understanding of the issue's nature. Whether identifying causes of invalidity or verifying valid reports, its rationales consistently reflect deep comprehension rather than shallow textual inference. Furthermore, we leveraged the additional key evidence collected by \appname in four cases to challenge maintainers’ original decisions, and all were ultimately revised, demonstrating the practical value of our approach.

\section{Data Availability Statement} 
\label{data-availability}
The replication package of this paper, including the complete prompts, source code, experimental results, etc., is publicly available with DOI: \url{https://doi.org/10.6084/m9.figshare.33993610}.

\section*{Acknowledgment}
This work is supported by the National Natural Science Foundation of China Grants (62572048 and 62232003).

\bibliographystyle{IEEEtran}
\balance
\bibliography{references}
\end{document}